\documentclass[aps,pra,twocolumn,superscriptaddress,showpacs]{revtex4-2}

\usepackage[utf8]{inputenc}

\usepackage{array}
\usepackage{amsmath}
\usepackage{bm}
\usepackage{graphicx}
\usepackage{epstopdf}
\usepackage{color}
\usepackage{amsfonts}
\usepackage{subfigure}
\usepackage{amssymb}
\usepackage{graphicx}
\usepackage{makecell}
\usepackage[colorlinks,citecolor=blue]{hyperref}
\usepackage{changes}
\newcommand{\onecolm}{
  \end{multicols}
  \vspace{-3.5ex}
  \noindent\rule{0.5\textwidth}{0.1ex}\rule{0.1ex}{2ex}\hfill
}
\newcommand{\twocolm}{
  \hfill\raisebox{-1.9ex}{\rule{0.1ex}{2ex}}\rule{0.5\textwidth}{0.1ex}
  \vspace{-4ex}
  \begin{multicols}{2}
}

\begin{document}

\title{Gauge-Invariant Non-Hermitian Quantum Theory: Foundation and Applications to Dynamical Phase Transitions}
\author{Fei Wang}
\affiliation{Institute of Biophysics, Dezhou University, Dezhou 253023, China}
\author{Guoying Liang}
\affiliation{Institute of Biophysics, Dezhou University, Dezhou 253023, China}
\author{Zecheng Zhao}
\affiliation{Institute of Biophysics, Dezhou University, Dezhou 253023, China}
\author{Bao-Ming Xu}
%\author{Li-Sha Guo}
\email{xubm2018@163.com}
\affiliation{Institute of Biophysics, Dezhou University, Dezhou 253023, China}
\date{Submitted \today}

\begin{abstract}
The description of states and dynamics in non-Hermitian systems is fundamentally linked to the choice of an appropriate theoretical framework---a point of ongoing debate in the field. This work addresses this issue by proposing a consistent formulation that reconciles existing controversies and establishes a unified theoretical understanding. Our approach rests on two foundational premises: (i) the dynamics of both left and right-vectors of a non-Hermitian system must satisfy the Schr\"{o}dinger equation; (ii) the theoretical framework must preserve gauge invariance, ensuring that physical quantities are independent of unobservable phase choices. Building on these physically motivated assumptions, we refine the biorthogonal framework, leading to a gauge-invariant non-Hermitian quantum theory. Our framework naturally encompasses the open-system effective non-Hermitian evolution as a special case, and can naturally reduce to standard quantum mechanics in the Hermitian limit. As a concrete application, we analyze the dynamical phase transition in a one-dimensional Su-Schrieffer-Heeger (SSH) model within this gauge-invariant non-Hermitian quantum theory. Notably, our formulation naturally generalizes the known condition for such transitions in Hermitian two-band systems, namely, $\mathbf{d}_{k}^i\cdot\mathbf{d}_{k}^f=0$, to the non-Hermitian case, where it takes the form $\mathrm{Re}\Bigl[\frac{\mathbf{d}_{k}^i}{d_{k}^i}\cdot\frac{\mathbf{d}_{k}^f}{d_{k}^f}\Bigr]=0$. Furthermore, we identify entirely new dynamical phase transitions that cannot be characterized by the winding number. We hope that this gauge-invariant non-Hermitian quantum theory will find broad applications in the study of non-Hermitian systems. 
\end{abstract}

\maketitle

\section{Introduction}\label{introduction}
In conventional quantum mechanics, the dynamics of a closed quantum system is fundamentally governed by a Hermitian Hamiltonian. The Hermiticity of operator ensures the real energy spectrum and the existence of a basis set of orthogonal eigenstates, meaning that an arbitrary state of the system can be expressed as a linear superposition of projections onto the subspaces spanned by this basis. The wide range of phenomena arising from this principle is well established and typically serves as the first introduction to the foundations of quantum mechanics. In reality, however, many systems are open or experience gain and loss---examples include photonic systems with losses \cite{R1}, open quantum systems subject to dissipation \cite{R2,R3}, and systems involving measurement and postselection \cite{R4,R5}. In such cases, the Hamiltonian generally ceases to be Hermitian.

The notion that the energy eigenvalues of an open system can effectively become complex dates back at least to 1928, when Gamow employed quantum theory to study nuclear radioactive decay \cite{R6}. Later, Lee and Yang introduced a complex magnetic field to construct a non-Hermitian Hamiltonian and mathematically analyzed ferromagnetic phase transitions using the zeros of the partition function \cite{R7,R8}. These Lee-Yang zeros are not merely mathematical constructs but have also been observed experimentally \cite{R9,R10}. Another significant contribution came from Bender et al., who proposed reformulating quantum theory by replacing the mathematical axiom of Hermiticity for the Hamiltonian and other observables with a more physically motivated condition tied to symmetry, such as $\mathcal{PT}$ symmetry \cite{R11,R12,R13}. This idea was further developed by Mostafazadeh \cite{R14,R15,R16}, who employed the concept of pseudo-Hermiticity to establish general conditions for a real spectrum. Although subsequent developments did not fully realize the original goal of constructing an alternative quantum framework, they led to the discovery of rich physical phenomena, including unconventional phase transitions \cite{R17,R18,R19,R20}, exceptional points \cite{R21,R22,R23,R24,R25,R26,R27}, the NH skin effect \cite{R28,R29,R30,R31,R32,R33,R34,R35,R36}, exotic supersonic modes in out-of-equilibrium systems \cite{R37,R38}, and novel topological phases \cite{R39,R40,R41,R42,R43,R44,R45,R46,R47}. Beyond theoretical advances, non-Hermitian quantum mechanics has been demonstrated experimentally on a variety of platforms, including photonic \cite{R48,R49,R50,R51,R52}, matter-light \cite{R53,R54,R55}, and electronic systems \cite{R20}, highlighting its broad relevance and potential for technological innovation.

A central issue in analyzing the physical properties of non-Hermitian systems concerns the choice of an appropriate framework to describe their states and dynamics \cite{R20,R22,R56}. A complete understanding of the spectral properties of non-Hermitian operators necessitates the simultaneous introduction of both left- and right-eigenvectors, forming the biorthogonal theoretical framework \cite{R57}. In practice, however, many studies consider only the right-eigenvectors. Even within the biorthogonal framework, a conceptual asymmetry arises: The dynamical evolution of the system is governed by the (non-Hermitian) Schr\"{o}dinger equation obeyed by the right-vectors, whereas the corresponding \textit{associated state} spanned by the left-eigenvectors does not satisfy the same dynamical equation \cite{R57,R58}. This asymmetry points to an incompleteness in the standard biorthogonal formulation--the duality of the left- and right-vectors is established for spectral decomposition, but not extended consistently to the dynamical level. 

A second, equally fundamental issue concerns the \textit{gauge structure} of non-Hermitian quantum mechanics. In conventional quantum mechanics, the global phase of a state vector carries no physical significance, and all observables are constructed to be manifestly gauge invariant. In the conventional biorthogonal framework, however, the \textit{associated state} is defined algebraically rather than dynamically, leading to a breakdown of gauge covariance. Consequently, physical quantities such as expectation values and transition probabilities acquire an explicit dependence on the phase choice.This ad hoc nature of defining the \textit{associated state} algebraically signals a deeper incompleteness in the dynamical description.

 Addressing these issues above and establishing a self-consistent dynamical description within the biorthogonal framework is the central goal of this work. Our approach is grounded in two fundamental principles: (i) the dynamics of both left- and right-vectors of a non-Hermitian system must satisfy the Schr\"{o}dinger equation, and (ii) equally essential, the framework must be manifestly gauge invariant, with physical quantities constructed from gauge-invariant combinations of left- and right-vectors. These principles are not independent: as we shall demonstrate, gauge invariance is restored precisely when both vectors are treated as dynamically independent entities satisfying their respective Schr\"{o}dinger equations. Based on these well-motivated premises, we revisit the concept of the \textit{associated state} within the biorthogonal framework, leading to a consistent reformulation of non-Hermitian quantum theory. This reformulation encompasses the representation of quantum states, the definition of expectation values for non-Hermitian operators, a revised quantum measurement theory, the dynamical evolution of states, and the corresponding geometric phases. 
 
 As a concrete application, we demonstrate the dynamic phase transition in a one-dimensional SSH model within this refined framework. Importantly, our formulation naturally generalizes the known conditions for the occurrence of such transitions of the Hermitian two-band systems to their non-Hermitian counterparts, thereby offering a unified perspective on a broader class of phenomena. Furthermore, some entirely new dynamical phase transitions that cannot be characterized by the winding number are discovered based on our formulation.

This paper is organized as follows: In the next section, we briefly review the traditional biorthogonal framework and clarify the points of contention within it. In Sec. \ref{Refined bio quantum mechanics}, we refine the biorthogonal framework to reconcile existing controversies and establish a gauge-invariant non-Hermitian quantum theory. Sec. \ref{Application in DQPT} applies this gauge-invariant non-Hermitian quantum theory to dynamical phase transition. Finally, Sec. \ref{Conclusion} closes the paper with some concluding remarks.

\section{A briefly review of biorthogonal quantum mechanics} \label{Bio quantum mechanics}
To facilitate the subsequent in-depth discussion, we begin with a brief review of the biorthogonal framework \cite{R57} and clarify the points of contention within it. A non-Hermitian system can be experimentally realized by adjusting a well controlled parameter or some parameters. The Hamiltonian of the non-Hermitian system is $H\neq H^\dag$, its right- and left-eigenvectors satisfy the eigen-equations:
\begin{equation}\label{LR}
  H|\epsilon_n\rangle=E_n|\epsilon_n\rangle,~ \langle\tilde{\epsilon}_n|H=\langle\tilde{\epsilon}_n|E_n,
\end{equation}
respectively. The eigenvalue $E_n$ can be either complex or real, depending on whether the $\mathcal{PT}$-symmetry is broken or the Hamiltonian is pseudo-Hermitian. The non-Hermitian nature of the Hamiltonian makes both the right- and left-eigenvectors not satisfy the orthogonality, i.e.,
\begin{equation}\label{}
\langle\epsilon_n|\epsilon_m\rangle\neq0~ \mathrm{and}~ \langle\tilde{\epsilon}_m|\tilde{\epsilon}_n\rangle\neq0,
\end{equation}
where $\langle\epsilon_n|=(|\epsilon_n\rangle)^\dag$ and $|\tilde{\epsilon}_n\rangle=(\langle\tilde{\epsilon}_n|)^\dag$. However, the right- and left-eigenvectors are orthogonal to each other:
\begin{equation}\label{}
  \langle\tilde{\epsilon}_m|\epsilon_n\rangle=\delta_{mn}.
\end{equation}
According to this orthogonality, the right- and left-eigenvectors should be normalized in the following way:
\begin{equation}\label{}
  |\epsilon_n\rangle\rightarrow\frac{1}{\sqrt{\langle\tilde{\epsilon}_n|\epsilon_n\rangle}}|\epsilon_n\rangle, ~
  \langle\tilde{\epsilon}_n|\rightarrow\frac{1}{\sqrt{\langle\tilde{\epsilon}_n|\epsilon_n\rangle}}\langle\tilde{\epsilon}_n|,
\end{equation}
when solving the right- and left-eigenvectors for a given non-Hermitian Hamiltonian. The completeness relation can be established
\begin{equation}\label{complete}
  \sum_n|\epsilon_n\rangle\langle\tilde{\epsilon}_n|=\mathbb{I}
\end{equation}
with $\mathbb{I}$ being the identity matrix.

For an arbitrary state 
\begin{equation}
|\psi\rangle=\sum_n c_n|\epsilon_n\rangle,
\end{equation}
an \textit{associated state} is defined: 
\begin{equation}\label{associated state}
\langle\bar{\psi}|=\sum_n\langle\tilde{\epsilon}_n| c^\ast_n, 
\end{equation}
where $c^\ast_n$ denotes the complex conjugate of $c_n$. Note that the \textit{associated state} is denoted by an overline, in contrast to the tilde used for left-eigenvectors.

Time evolution is governed by the Hamiltonian, with the evolution operator given by (setting $\hbar=1$)
\begin{equation}
U(t)=e^{-iHt}, 
\end{equation}
which is non-unitary due to the non-Hermiticity of $H$. At time $t$, the state becomes 
\begin{equation}\label{NHE}
|\psi(t)\rangle=e^{-iHt}|\psi\rangle=\sum_n c_n e^{-iE_n t}|\epsilon_n\rangle. 
\end{equation}
Its \textit{associated state} evolves accordingly:
\begin{equation}\label{associated state t}
\langle\bar{\psi}(t)|=\sum_n\langle\tilde{\epsilon}_n| e^{iE^\ast_n t}c^\ast_n. 
\end{equation}

The definition of the \textit{associated state} reveals a subtle tension with the \textit{gauge structure} of non-Hermitian quantum mechanics. Under a gauge transformation $|\psi(t)\rangle\rightarrow|\psi'(t)\rangle=e^{i\phi(t)}|\psi(t)\rangle$, the coefficients transform as $c_n\rightarrow e^{i\phi(t)}c_n$. Because the \textit{associated state} is constructed by complex conjugation [Eq.~\eqref{associated state}], its formal structure requires the transformation $\langle\bar{\psi}(t)|\rightarrow\langle\bar{\psi}(t)|e^{-i\phi^\ast(t)}$. For non-Hermitian systems, however, the phase $\phi(t)$ is generically complex due to the non-unitary evolution, so that $\phi^\ast(t)\neq\phi(t)$. The expectation value of an observable $F$ therefore acquires a phase-dependent prefactor: $\langle\bar{\psi}'(t)|F|\psi'(t)\rangle=e^{-2\,\mathrm{Im}\,\phi(t)}\langle\bar{\psi}(t)|F|\psi(t)\rangle$, which cannot be eliminated by any local redefinition of the phase. This indicates that gauge covariance is not preserved within the 
conventional framework.

To address this, the conventional framework introduces normalizing denominators, defining the expectation value of $F$ as
\begin{equation}\label{expec}
\langle F\rangle=\frac{\langle\bar{\psi}|F|\psi\rangle}{\langle\bar{\psi}|\psi\rangle}.
\end{equation}
While this ratio formally cancels the phase-dependent prefactor, the procedure raises conceptual questions. The denominator $\langle\bar{\psi}|\psi\rangle$ serves as a compensating factor rather than a physical invariant. The origin of this complication lies in the construction of the left-vector: by complex conjugation rather than dynamical evolution, it is treated as an auxiliary entity rather than as part of a unified dynamical description.

Equation~\eqref{LR} is the stationary Schr\"odinger equation for the non-Hermitian Hamiltonian $H$, where both right- and left-eigenvectors satisfy the same eigenvalue equation governed by $H$. This structure suggests that, in dynamics, both vectors should participate on equal footing. Indeed, in standard quantum mechanics, the time-dependent Schr\"odinger equation for the right-vector automatically implies the conjugate equation for the left-vector because the two are related by Hermitian conjugation. In other words, the Hermitian conjugate construction of the left-vector commonly adopted in Hermitian quantum mechanics is valid precisely because such a construction guarantees that the left-vector satisfies the Schr\"odinger equation. In non-Hermitian systems, however, complex conjugation no longer preserves this dynamical consistency: the \textit{associated state} defined in Eq.~\eqref{associated state} does not satisfy the Schr\"odinger equation,
\begin{equation}\label{associated state t1}
-i\frac{\partial}{\partial t}\langle\bar{\psi}(t)|\neq\langle\bar{\psi}(t)|H.
\end{equation}
It is perhaps for this reason that many studies on non-Hermitian systems have abandoned the use of the \textit{associated state} altogether, opting instead to directly normalize the right-eigenvectors within their formulations. Yet this approach sacrifices the biorthogonal structure that is essential for a complete description of non-Hermitian spectra.

To reconcile these issues, we propose in the next section a refined formulation that restores the dynamical unity of left- and right-vectors. Our approach rests on a foundational premise: the dynamics of both left- and right-vectors of a non-Hermitian system must satisfy the Schr\"odinger equation. As we shall demonstrate, this requirement is not merely a matter of theoretical elegance; it is demanded by the \textit{gauge structure} of quantum mechanics and leads naturally to a consistent, gauge-invariant framework.

\section{Gauge-Invariant Non-Hermitian Quantum Theory} \label{Refined bio quantum mechanics}
\textit{State.--- }
According to the completeness relation of Eq. \eqref{complete}, the non-Hermitian Hamiltonian can be expressed as
\begin{equation}\label{spectrum}
  H=\sum_n E_n|\epsilon_n\rangle\langle\tilde{\epsilon}_n|, 
\end{equation}
which constitutes its spectral decomposition. Importantly, for a non-Hermitian system $H$, the quantum state corresponding to an energy $E_n$ must be described jointly by both the left- and right-eigenvectors, not by the right-eigenvector alone--unlike in Hermitian systems. In Hermitian systems, the left-eigenvector is simply the conjugate transpose of the right-eigenvector, so that the state is fully captured by the right-eigenvector. In non-Hermitian systems, however, no such simple relation exists; the left-eigenvector cannot be fully determined from the right-eigenvector alone. In other words, for non-Hermitian systems, an experimentally observed state must be treated as a unified description of both the left- and right-eigenvectors, which together form an inseparable entity. In view of this intrinsic duality, we adopt the density matrix $|\epsilon_n\rangle\langle\tilde{\epsilon}_n|$ to describe the eigenstate of the non-Hermitian Hamiltonian $H$ with energy $E_n$. Using this density matrix the phase transitions in the non-Hermitian XY model has been investigated in Ref. \cite{Xu2026}

An arbitrary pure state can be expressed as $\rho=|\psi\rangle\langle\tilde{\psi}|$ satisfying $\langle\tilde{\psi}|\psi\rangle=1$. It can be spanned by the left- and right-eigenvectors of $H$:
\begin{equation}\label{}
  \rho=|\psi\rangle \langle\tilde{\psi}|=\sum_{mn}c_m\tilde{c}_n|\epsilon_m\rangle\langle\tilde{\epsilon}_n|,
\end{equation}
i.e.,
\begin{equation}\label{state}
  |\psi\rangle=\sum_nc_n|\epsilon_n\rangle, ~
  \langle\tilde{\psi}|=\sum_n\langle\tilde{\epsilon}_n|\tilde{c}_n
\end{equation}
with
\begin{equation}\label{cn}
  c_n=\langle\tilde{\epsilon}_n|\psi\rangle,~ \tilde{c}_n=\langle\tilde{\psi}|\epsilon_n\rangle.
\end{equation}
It should be noted that $\tilde{c}_n\neq(c_n)^\ast$ in general. Here, the left-vector $\langle\tilde{\psi}|$ and its associated coefficient $\tilde{c}_n$ are  denoted with a tilde to distinguish them from the \textit{associated state} in the conventional biorthogonal framework (where the \textit{associated state} is marked with an overline). This choice reflects the physical consistency of $\langle\tilde{\psi}|$ with the left-eigenvector $\langle\tilde{\epsilon}_n|$ of the non-Hermitian Hamiltonian, a point that will be elaborated on later.

The relation $\langle\tilde{\psi}|\psi\rangle=1$ means that $\sum_nc_n\tilde{c}_n=1$. Note that the diagonal element $\rho_{nn}(t)=c_n\tilde{c}_n$, i.e., the probability of energy level $|\epsilon_n\rangle\langle\tilde{\epsilon}_n|$, is not always positive, but can be negative and even complex. Like complex energy, we argue that complex probability is a remarkable feature of non-Hermitian quantum mechanics. This feature is not unprecedented in quantum theory. In 1932, Wigner introduced the first quasiprobability distribution to describe quantum systems in phase space, demonstrating that quantum mechanics permits probability-like quantities that violate the non-negativity constraint of classical probability theory \cite{Wigner}. Subsequent developments, including the Husimi $Q$-distribution \cite{Husimi} and the Glauber-Sudarshan $P$-distribution \cite{Glauber,Sudarshan}, further established that quasiprobabilities are indispensable tools in quantum optics. On the other hand, as early as 1933, Kirkwood introduced a complex-valued quasiprobability distribution for quantum states in phase space \cite{Kirkwood}, and Dirac independently developed similar ideas in his transformation theory in 1945 \cite{Dirac}. The Kirkwood-Dirac quasiprobability arises from the impossibility of jointly diagonalizing noncommuting observables. Its complex phase encodes the relative orientation between incompatible bases, a structure that persists independently of operator ordering. This demonstrates that complex probability-like quantities are intrinsic to quantum mechanics whenever noncommuting observables are jointly represented. This complex nature, long regarded as a mathematical curiosity, has found renewed relevance in quantum thermodynamics \cite{Pei2023,Xu2018,Yoshimura2026,Lostaglio2023,Gherardini2024}, quantum many-body physics \cite{Gherardini2024,Halpern2017,Halpern2018,Alonso2019,Shukla2026}, and fluctuation theorems \cite{Levy2020,Zhang2024,Lostaglio2018,Zhang2026}. Crucially, Kirkwood-Dirac quasiprobabilities have been experimentally reconstructed \cite{Hernandez2024a,Hernandez2024b,Lupu2022,Solinas2021}, demonstrating that complex-valued probability-like quantities are empirically accessible objects. The emerging complex probabilities suggest a fundamental extension of probability theory beyond the Kolmogorov axioms, a development already underway in mathematics \cite{R59}. This progression compels quantum mechanics itself to embrace complex probabilities as fundamental physical quantities.

Given an  arbitrary right-vector $|\psi\rangle$, how does one determine its associated left-vector $\langle\tilde{\psi}|$ satisfying Eqs.~\eqref{state} and \eqref{cn}? Or conversely, given an arbitrary left-vector $\langle\tilde{\psi}|$, how does one determine its associated right-vector $|\psi\rangle$? In the Hermitian systems, the corresponding left-vector that satisfies $\langle\psi|\psi\rangle=1$ can be completely determined through transposed conjugation: $\langle\psi|=(|\psi\rangle)^\dag$. However, in the non-Hermitian systems, the situation is considerably more complicated. For a given right-vector $|\psi\rangle=\sum_nc_n|\epsilon_n\rangle$, it was previously believed that the \textit{associated state} $\langle\bar{\psi}|\psi\rangle=1$ is constructed by $\langle\bar{\psi}|=\sum_n\langle\tilde{\epsilon}_n|c_n^\ast$ \cite{R57}. This belief, however, contradicts Schr\"{o}dinger equation. Mathematically, there exists infinitely many associated left-vector for a given right-vector that satisfing $\langle\bar{\psi}|\psi\rangle=1$. To illustrate this in a two-dimensional space, consider the right-vector $|\psi\rangle=1/2(|1\rangle+\sqrt{2}|0\rangle)$. We can define two associated left-vectors: $\langle\tilde{\psi}_1|=\langle1|+\sqrt{2}/2\langle0|$ and $\langle\tilde{\psi}_2|=1/2\langle1|+3\sqrt{2}/4\langle0|$. Their linear combinations $\langle\tilde{\psi}|=p\langle\tilde{\psi}_1|+(1-p)\langle\tilde{\psi}_2|$ with $p\in[0,1]$ also yield associated left-vectors. Despite all these states being associated with $|\psi\rangle$, they possess fundamentally different physical natures, as evidenced by the fact that $|\psi\rangle\langle\tilde{\psi}_1|$ and $|\psi\rangle\langle\tilde{\psi}_2|$ are respectively the eigenstates of non-commuting physical operators $\mathcal{O}_1=|1\rangle\langle0|+2|0\rangle\langle1|$ and $\mathcal{O}_2=1/2|1\rangle\langle1|-3\sqrt{2}/4|1\rangle\langle0|-\sqrt{2}/2|0\rangle\langle1|-1/2|0\rangle\langle0|$. This highlights the rich structure of non-Hermitian systems, where multiple left-vectors can correspond to a single right-vector, and vice versa, each with distinct physical implications. Therefore, knowing only a right- or left-vector does not fully encapsulate the underlying physics. 

We argue that for a meaningful physical interpretation, an arbitrary pure state $|\psi\rangle\langle\tilde{\psi}|$ must be regarded as an inseparable entity. In particular, it must be representable as a rotation of a given eigenstate of the non-Hermitian system in the state space, i.e., $|\psi\rangle\langle\tilde{\psi}|=U|\epsilon_n\rangle\langle\tilde{\epsilon}_n|U^{-1}$, where $U$ is a rotation or transformation operator. Importantly, in non-Hermitian quantum mechanics, $U$ is not necessarily unitary. In other words, $|\psi\rangle\langle\tilde{\psi}|$ must be an eigenstate of a physical operator $\mathcal{O}=UHU^{-1}$. This requirement has physical significance, as a pure state is necessarily an eigenstate of some operator. In this sense, $|\psi\rangle$ and $\langle\tilde{\psi}|$ cannot be chosen independently; rather, they must be determined simultaneously, either by diagonalizing the physical operator under consideration or by applying a suitable transformation to a given state. Only in this way can the Schr\"{o}dinger equation be satisfied for both right- and left-vectors, which will be discussed in the following.

\textit{Dynamics.---}
Given a non-Hermitian Hamiltonian $H$ and an initial state $\rho(0)=|\psi(0)\rangle\langle\tilde{\psi}(0)|$, both the left- 
and right-vectors of the system satisfy the Schr\"odinger equation ($\hbar=1$):
\begin{equation}\label{Schrodinger equation}
    i\frac{\partial}{\partial t}|\psi(t)\rangle = H|\psi(t)\rangle, \quad
    -i\frac{\partial}{\partial t}\langle\tilde{\psi}(t)| = \langle\tilde{\psi}(t)|H.
\end{equation}
Equivalently, the density matrix $\rho(t)=|\psi(t)\rangle\langle\tilde{\psi}(t)|$ satisfies the Liouville-von Neumann equation
\begin{equation}\label{Liouville}
    \frac{d\rho(t)}{dt} = -i[H,\rho(t)].
\end{equation}
The solution takes the form
\begin{equation}\label{}
    \rho(t) = e^{-iHt}|\psi(0)\rangle\langle\tilde{\psi}(0)|e^{iHt},
\end{equation}
which corresponds to a non-unitary rotation of the initial state. This evolution, non-unitary because $H \neq H^\dagger$, was adopted in Refs.~\cite{R60,R61,R62}.

From Eq.~(\ref{state}), the time-dependent coefficients are
\begin{equation}\label{}
    |\psi(t)\rangle = \sum_n e^{-i\epsilon_n t}c_n|\epsilon_n\rangle = \sum_n c_n(t)|\epsilon_n\rangle,
\end{equation}
\begin{equation}\label{}
    \langle\tilde{\psi}(t)| = \sum_n \langle\tilde{\epsilon}_n|\tilde{c}_n e^{i\epsilon_n t} = \sum_n \langle\tilde{\epsilon}_n|\tilde{c}_n(t),
\end{equation}
yielding the density matrix
\begin{equation}\label{rhot}
    \rho(t) = \sum_{mn} c_m(t)\tilde{c}_n(t)|\epsilon_m\rangle\langle\tilde{\epsilon}_n|.
\end{equation}

The diagonal elements $\rho_{nn}(t) = c_n(t)\tilde{c}_n(t)$, interpretable as the probability of the state $|\epsilon_n\rangle\langle\tilde{\epsilon}_n|$, are not restricted to positive real values but may be negative or complex. This is not a defect of the formalism but a physical feature: under a gauge transformation $|\psi(t)\rangle \to e^{i\phi(t)}|\psi(t)\rangle$, $\langle\tilde{\psi(t)}| \to \langle\tilde{\psi}(t)|e^{-i\phi(t)}$, the density matrix $\rho(t)$ is strictly invariant, and consequently its trace is preserved. The normalization condition $\sum_n c_n(t)\tilde{c}_n(t) = 1$ is thus not imposed externally but emerges as the gauge-invariant residue of the theory---the only combination of complex probabilities immune to arbitrary phase redefinitions.

Alternatively, the dynamics of non-Hermitian systems are sometimes formulated in a form of $\dot{\rho}=-i[H_{eff}\rho-\rho H_{eff}^{\dag}]$, typically derived from a quantum master equation by neglecting environmental jump terms \cite{R63}. While this formulation preserves the Hermiticity of the density matrix of the system, it does not possess gauge-invariance. The refined dynamics \eqref{Schrodinger equation} or \eqref{Liouville} encompasses this formulation as special cases. If one only focus the right-vector and imposes the conjugate constraint $\langle\tilde{\psi}(t)| \equiv \langle\psi(t)| = (|\psi(t)\rangle)^\dagger$, the density matrix becomes Hermitian, $\varrho(t) = |\psi(t)\rangle\langle\psi(t)|$, 
and Eq.~\eqref{Liouville} reduces to
\begin{equation}
\frac{d\varrho(t)}{dt} = -i[H\varrho(t) - \varrho(t) H^\dagger],
\end{equation}
which is precisely the effective non-Hermitian evolution derived from quantum master equations by neglecting environmental jump terms \cite{R63}. Thus, the open-system effective description emerges as a gauge-fixed subspace of our framework, valid when the left-vector is strictly tied to the right-vector by Hermitian conjugation. 

\textit{Measurement.---}
Given an operator $A$, its expectation value on the density matrix $\rho$ is
\begin{equation}\label{}
    \langle A\rangle = \mathrm{Tr}[A\rho] = \sum_n \langle\tilde{\epsilon}_n|A\rho|\epsilon_n\rangle.
\end{equation}
For the spectral decomposition $A = \sum_n a_n|a_n\rangle\langle\tilde{a}_n|$, 
the projective measurement operator for outcome $a_n$ is
\begin{equation}\label{}
    M_n = |a_n\rangle\langle\tilde{a}_n|,
\end{equation}
satisfying the completeness relation $\sum_n M_n = \mathbb{I}$. 
The probability of obtaining $a_n$ is
\begin{equation}\label{}
    p_n = \mathrm{Tr}[M_n\rho] = \langle\tilde{a}_n|\rho|a_n\rangle,
\end{equation}
and the post-measurement state is
\begin{equation}\label{}
    \frac{M_n\rho M_n}{\mathrm{Tr}[M_n\rho M_n]} = |a_n\rangle\langle\tilde{a}_n|.
\end{equation}
In particular, the probability $p_n$ may be negative or complex, 
a distinctive feature of non-Hermitian quantum mechanics. If the 
measurement result is disregarded, the post-measurement state 
becomes
\begin{equation}\label{}
    \sum_n \frac{M_n\rho M_n}{\mathrm{Tr}[\sum_n M_n\rho M_n]} = \sum_n p_n |a_n\rangle\langle\tilde{a}_n|.
\end{equation}

The gauge invariance of the formalism is manifest. Under a gauge transformation $|\psi\rangle \to e^{i\phi}|\psi\rangle$, 
$\langle\tilde{\psi}| \to \langle\tilde{\psi}|e^{-i\phi}$, the density matrix $\rho$ is strictly invariant. Consequently, the expectation value of $A$, remains unchanged, as does the probability $p_n = \mathrm{Tr}[M_n\rho]$ and the post-measurement state. This is because the measurement operators $M_n = |a_n\rangle\langle\tilde{a}_n|$ transform covariantly: under the gauge rotation of the biorthogonal basis, $|a_n\rangle \to e^{i\phi}|a_n\rangle$ and $\langle\tilde{a}_n| \to \langle\tilde{a}_n|e^{-i\phi}$, leaving $M_n$ invariant. Thus, gauge invariance is not restored post hoc through compensating denominators, but is intrinsic to the framework: the trace operation $\mathrm{Tr}[A\rho]$ automatically preserves gauge-invariance, thereby ensuring that all physical predictions are independent of unobservable phase choices.

\textit{Loschmit echo and geometric phase.---} 
The Loschmit echo or return probability at time $t$ is
\begin{equation}\label{}
  \mathcal{L}(t)=\langle\tilde{\psi}(0)|\psi(t)\rangle\langle\tilde{\psi}(t)|\psi(0)\rangle.
\end{equation}
We can define a complex total phase
\begin{equation}\label{}
  \Phi_{tot}(t)=-i\log\frac{\langle\tilde{\psi}(0)|\psi(t)\rangle}{\sqrt{\mathcal{L}(t)}},
\end{equation}
making $\langle\tilde{\psi}(0)|\psi(t)\rangle=\sqrt{\mathcal{L}(t)}e^{i\Phi_{tot}(t)}$ and $\langle\tilde{\psi}(t)|\psi(0)\rangle=\sqrt{\mathcal{L}(t)}e^{-i\Phi_{tot}(t)}$. Under
a gauge transformation $|\psi(t)\rangle\rightarrow|\psi'(t)\rangle=e^{i\phi(t)}|\psi(t)\rangle$ and $\langle\tilde{\psi}(t)|\rightarrow\langle\tilde{\psi}'(t)|=\langle\tilde{\psi}(t)|e^{-i\phi(t)}$, we have that $\Phi_{tot}(t)\rightarrow\Phi'_{tot}(t)=\Phi_{tot}(t)+\phi(t)-\phi(0)$ and $-i\langle\tilde{\psi}(t)|\dot{\psi}(t)\rangle=-\langle\tilde{\psi}(t)|H(t)|\psi(t)\rangle
\rightarrow-i\langle\tilde{\psi}'(t)|\dot{\psi}'(t)\rangle=-\langle\tilde{\psi}(t)|H(t)|\psi(t)\rangle+\dot{\phi}(t)$. The Hamiltonian here has been changed to time-dependent form, in order to discuss more general cases. From these properties we can construct the Pancharatnam geometric phase, which is gauge-invariant:
\begin{equation}\label{}
  \Phi_g(t)=\Phi_{tot}(t)-\Phi_{d}(t)
\end{equation}
with
\begin{equation}\label{}
\begin{split}
  \Phi_{d}(t)&=-i\int_0^t\langle\tilde{\psi}(t')|\dot{\psi}(t')\rangle dt'\\
  &=-\int_0^t\langle\tilde{\psi}(t')|H(t')|\psi(t')\rangle dt'
\end{split}
\end{equation}
being the dynamical phase. These dynamical and geometric phases are exactly those proposed in Refs. \cite{R61,R62}. If the Hamiltonian is restricted to change very slowly, satisfying the adiabatic condition
\begin{equation}\label{}
  \Bigg|\frac{\langle\tilde{\varepsilon}_m(t)|\dot{H}(t)|\varepsilon_n(t)\rangle}{E_m(t)-E_n(t)}\Bigg|\ll1,~ \mathrm{for}~ \mathrm{all}~ m\neq n,
\end{equation}
the dynamical and geometric phases will become
\begin{equation}\label{}
  \phi_{n}(t)=\int_0^tE_n(t')dt'
\end{equation}
and
\begin{equation}\label{}
  \gamma_{n}(t)=i\int_0^t\langle\tilde{\varepsilon}_n(t')|\frac{d}{dt'}|\varepsilon_n(t')\rangle dt',
\end{equation}
which recovers the Berry's insights in non-Hermitian quantum mechanics.

\section{Applications of Non-Hermitian Quantum Mechanics in Dynamical Phase Transition} \label{Application in DQPT}
In this section, we use this gauge-invariant non-Hermitian quantum theory to investigate DQPT in non-Hermitian systems. First, we repeat Heyl's route of establishing DQPT in Hermitian systems and extend it to non-Hermitian systems. In the equilibrium phase transition of Hermitian system, the fundamental quantity of interest is the canonical partition function. Here, we turn our attention on the non-Hermitian systems and define
\begin{equation}\label{}
  Z(\beta)=\mathrm{Tr}[e^{-\beta H}]=\sum_{n}\langle\tilde{\epsilon}_n|e^{-\beta H}|\epsilon_n\rangle=\sum_ne^{-\beta \epsilon_n}
\end{equation}
with $\beta$ being the inverse temperature. The boundary partition function with the boundary state $|\psi_i\rangle\langle\tilde{\psi}_i|$ is
\begin{equation}\label{boundary partition function}
  \mathcal{Z}(\beta)=\mathrm{Tr}[e^{-\beta H}|\psi_i\rangle\langle\tilde{\psi}_i|]=\langle\tilde{\psi}_i|e^{-\beta H}|\psi_i\rangle.
\end{equation}
This quantity is similar to Loschmidt amplitude of the initial state $|\psi_i\rangle\langle\tilde{\psi}_i|$ after time evolution
\begin{equation}\label{Loschmidt}
  \mathcal{G}(t)=\langle\tilde{\psi}_i|e^{-iHt}|\psi_i\rangle.
\end{equation}
Due to this similarity, we expend time $t$ and the temperature $\beta$ into the complex plane and focus on the boundary partition function
\begin{equation}\label{Zz}
  \mathcal{Z}(z)=\langle\tilde{\psi}_i|e^{-zH}|\psi_i\rangle,
\end{equation}
where $z\in \mathbb{C}$. For imaginary $z=it$ this can be understood as the probability amplitude in the non-Hermitian realm. For real $z=\beta$ it just describes the overlap amplitude of Eq. (\ref{boundary partition function}). In the thermodynamic limit, the free energy density in the non-Hermitian realm can be defined as
\begin{equation}\label{free energy density}
  f(z)=-\lim_{N\rightarrow\infty}\frac{1}{N}\ln\mathcal{Z}(z),
\end{equation}
where $N$ is the number of degrees of freedom. Since the boundary partition function Eq. \eqref{Zz} is an entire function of $z$, according to the Weierstrass factorization theorem \cite{R64}, it can be expanded by its zeros $z_j\in C$:
\begin{equation}\label{}
  \mathcal{Z}(z)=e^{h(z)}\prod_j(1-\frac{z}{z_j}),
\end{equation}
where $h(z)$ is an entire function. Thus,
\begin{equation}\label{}
  f(z)=-\lim_{N\rightarrow\infty}\frac{1}{N}\biggl[h(z)+\sum_j\ln\biggl(1-\frac{z}{z_j}\biggr)\biggr]
\end{equation}
and the nonanalytic part of the free energy density is solely determined by the zeros $z_j$. This concept was originally made by M. E. Fisher to discuss the temperature-driven phase transition \cite{R65}. This observation is analogous to the Lee-Yang analysis of equilibrium phase transitions in the complex magnetic field plane \cite{R7,R8}. An equilibrium  phase transition in non-Hermitian systems for nonanalytic behavior at a critical temperature temperature ($\beta_c$) would occur when such zeros are real. On the contrary, if such zeros are purely imaginary, a dynamical phase transition for nonanalytic behavior in time will occur, that is the breakdown of a short time expansion in the thermodynamic limit at a critical time $t_c$ \cite{R66,R67}. The dynamical phase transition means that the return probability (Loschmidt echo) at critical time $t_c$ vanishes, i.e,
\begin{equation}\label{}
\mathcal{L}(t_c)=\langle\tilde{\psi}_i|e^{-iHt_c}|\psi_i\rangle\langle\tilde{\psi}_i|e^{iHt_c}|\psi_i\rangle=0.
\end{equation}
A vanishing value of the Loschmidt echo, results in nonanalyticities of the rate function
\begin{equation}\label{}
  r(t)=-\lim_{N\rightarrow\infty}\frac{1}{N}\ln\mathcal{L}(t)=f(it)+f(-it),
\end{equation}
which is generally used to mark emergence of dynamical phase transition. In non-Hermitian quantum mechanics, the vanishing of the Loschmidt echo corresponds to the vanishing of its vector length in the complex plane, leading to non-analytic behavior in the real part of the associated rate function. Accordingly, our analysis of DQPT is confined to the real part of the Loschmidt echo rate function $\mathrm{Re}r(t)$.

\begin{figure}
\begin{center}
\includegraphics[width=8cm]{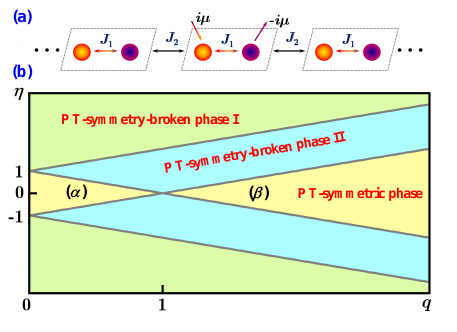}
\caption{(Color online) (a) The schematic diagram of SSH model. (b) The $\mathcal{PT}$-symmetry phase diagram in non-Hermitian SSH model. In the $\mathcal{PT}$-symmetric phase (yellow areas), the system energy is real. The $\mathcal{PT}$-symmetry-broken phase can be divided into two types: Phase I (green areas), characterized by a purely imaginary energy spectrum, and Phase II (blue areas), which exhibits a complex energy spectrum. In the phase diagram, $q=J_2/J_1$ and $\eta=\mu/J_1$. }
\label{figure1}
\end{center}
\end{figure}

We now work out these analytic properties explicitly for the one-dimensional Su-Schrieffer-Heeger (SSH) model \cite{R68,R69} (Fig. \ref{figure1}(a))
\begin{widetext}
\begin{equation}\label{}
  H=\sum_nJ_1(c_{n,A}^\dag c_{n,B}+c_{n,B}^\dag c_{n,A})+J_2(c_{n+1,A}^\dag c_{n,B}+c_{n,B}^\dag c_{n+1,A})
  +i\mu c_{n,A}^\dag c_{n,A}-i\mu c_{n,B}^\dag c_{n,B},
\end{equation}
\end{widetext}
where $c_{n,\gamma}^\dag$ and $c_{n,\gamma}$ are the creation and annihilation operators on sublattice $\gamma\in \{A, B\}$ on the $n$th unit cell. $\pm i\mu$ are the physical gain and loss. $J_1$ and $J_2$ are the transition amplitudes of the intracell and intercell hopping processes, respectively. In this paper, we set $J_1=1$ as the overall energy scale without loss of generality, and let $q=J_2/J_1$ and $\eta=\mu/J_1$ for simplicity. This Hamiltonian is $\mathcal{PT}$-symmetric, satisfying $[\mathcal{PT},H]=0$.

Such system exhibit a compact representation of the Hamiltonian
\begin{equation}\label{Hamiltonian}
\hat{H}=\sum_{k\in BZ}
\begin{pmatrix} c^\dag_{k,A} & c^\dag_{k,B} \end{pmatrix}
\begin{pmatrix} i\eta &1+qe^{ik} \\ 1+qe^{-ik} & -i\eta\end{pmatrix}
\begin{pmatrix} c_{k,A} \\ c_{k,B} \end{pmatrix},
\end{equation}
with the momentum summation extending over the first Brillouin zone (BZ). Here, $c_{k,A}$ and $c_{k,B}$ denote the spinors on sublattice $A$ and $B$. Each $H_k$ acts on a two-dimensional Hilbert space generated by \{$|10\rangle$,$|01\rangle$\}
where $c_{k,A}^\dag c_{k,B}|01\rangle=|10\rangle$, and can be represented in that basis by a $2\times2$ matrix
\begin{equation}\label{}
  H_k=\mathbf{d}_k\cdot\bm{\sigma},
\end{equation}
where 
\begin{equation}\label{}
  \mathbf{d}_k=(1+q\cos k,-q\sin k,i\eta),
\end{equation}
and $\bm{\sigma}$ the vector of standard Pauli matrices. The energy spectrum is then explicitly given by $\epsilon_k^\pm=\pm d_k$ with 
\begin{equation}\label{}
  d_k=\sqrt{|1+qe^{ik}|^2-\eta^2}.
\end{equation}
The corresponding right- and left-eigenvectors are
\begin{equation}\label{}
\begin{split}
|\epsilon^+_{k}\rangle&=\frac{1}{\sqrt{2\bigl(1+\frac{d_k^z}{d_k}\bigr)}}
\biggl[\Bigl(1+\frac{d_k^z}{d_k}\Bigr)|1\rangle+\Bigl(\frac{d_k^x}{d_k}+i\frac{d_k^y}{d_k}\Bigr)|0\rangle\biggr], \\
\langle\tilde{\epsilon}^+_{k}|&=\frac{1}{\sqrt{2\bigl(1+\frac{d_k^z}{d_k}\bigr)}}
\biggl[\Bigl(1+\frac{d_k^z}{d_k}\Bigr)\langle1|+\Bigl(\frac{d_k^x}{d_k}-i\frac{d_k^y}{d_k}\Bigr)\langle0|\biggr],
\end{split}
\end{equation}
and
\begin{equation}\label{}
\begin{split}
|\epsilon^-_{k}\rangle&=\frac{1}{\sqrt{2\bigl(1+\frac{d_k^z}{d_k}\bigr)}}
\biggl[\Bigl(-\frac{d_k^x}{d_k}+i\frac{d_k^y}{d_k}\Bigr)|1\rangle+\Bigl(1+\frac{d_k^z}{d_k}\Bigr)|0\rangle\biggr], \\
\langle\tilde{\epsilon}^-_{k}|&=\frac{1}{\sqrt{2\bigl(1+\frac{d_k^z}{d_k}\bigr)}}
\biggl[\Bigl(-\frac{d_k^x}{d_k}-i\frac{d_k^y}{d_k}\Bigr)\langle1|+\Bigl(1+\frac{d_k^z}{d_k}\Bigr)\langle0|\biggr],
\end{split}
\end{equation}
respectively.

If $|\eta|<|1-q|$, $\epsilon_k^\pm$ is real for all $k$, meaning the eigenstate of the system obey $\mathcal{PT}$-symmetry: $[\mathcal{PT},|\epsilon_k^\pm\rangle\langle\tilde{\epsilon}_k^\pm|]=0$. In the remaining parameter regime, the system enters the $\mathcal{PT}$-symmetry-broken phase with the complex energy, the modes satisfying $|\eta|>|1+qe^{ik}|$ have purely imaginary energies, whereas those with $|\eta|<|1+qe^{ik}|$ possess real energies. Along the boundary where $|\eta|=|1+qe^{ik}|$, the spectrum becomes degenerate, corresponding to the exceptional point. Notably, if $|\eta|>|1+q|$, the entire energy spectrum of the system is purely imaginary. Accordingly, we separate the $\mathcal{PT}$-symmetry-broken phase into two types: Phase I, characterized by a purely imaginary energy spectrum, and Phase II, which exhibits a complex energy spectrum. Based on these criteria, the phase diagram of non-Hermitian SSH model are shown in Fig. \ref{figure1}(b).

In a quantum quench experiment the system is prepared, at time $t<0$ in the ground state for parameter $\mathbf{d}_k^i$,
\begin{equation}\label{}
  |\psi_i\rangle\langle\tilde{\psi}_i|=\bigotimes_{k\in BZ}|\epsilon^-_{k}\rangle\langle\tilde{\epsilon}^-_{k}|.
\end{equation}
Then at time $t=0$ the parameter is suddenly changed from $\mathbf{d}_k^i$ to $\mathbf{d}_k^f$. This process is assumed to be sufficiently sudden that the state of the system remains unchanged initially. After quenching, the dynamics of the system is governed by the post-quench Hamiltonian. For such quench dynamics, we demonstrate that the conditions for the occurrence of a dynamical quantum phase transition are that the critical mode energy $d_{k_c}^f$ after quench to be real and 
\begin{equation}\label{condition}
  \mathrm{Re}\Biggl[\frac{\mathbf{d}_{k_c}^i}{d_{k_c}^i}\cdot\frac{\mathbf{d}_{k_c}^f}{d_{k_c}^f}\Biggr]=0.
\end{equation}
The detailed derivation of these conditions is provided in the appendix. The DQPT condition in Hermitian systems, $\mathbf{d}_{k_c}^i\cdot\mathbf{d}_{k_c}^f=0$, emerges as a special case of Eq. \eqref{condition}. More importantly, the orthogonality condition Eq. \eqref{condition} required for dynamic phase transitions can only be derived within the refined biorthogonal framework proposed by this work. In contrast, the conventional biorthogonal approach--or one that considers only the right eigenvectors--fails to yield a universal condition applicable to two-band systems, as the derivation in these frameworks depends on specific parametric details. Since $d^f_{k_c}$ should be real for DQPT, this paper, for simplicity, only considers the case that the post-quench system remains in the $\mathcal{PT}$-symmetric phase. 

\begin{figure}
\begin{center}
\includegraphics[width=8cm]{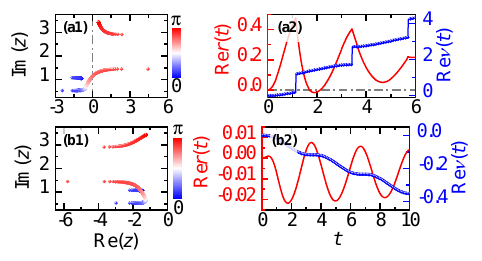}
\caption{(Color online) Lines of Fisher zeros (a1-b1), the time evolution of the real parts of the rate function $\mathrm{Re}r(t)$ [red curves in (a2) and (b2)] and the winding number $\mathrm{Re}\nu(t)$ [blue curves in (b1) and (b2)] for quenches $q=0.5\rightarrow q'=2$ (a1-a2) and $q=1.5\rightarrow q'=2$ (b1-b2). The other parameters are $\eta=\eta'=0.4$, and $l=0$.}
\label{figure2}
\end{center}
\end{figure}

\subsection{DQPT for the quench with the $\mathcal{PT}$-symmetric phase}
In this section, We consider the system is quenched within the $\mathcal{PT}$-symmetric phase. In this case, the condition of DQPT Eq. \eqref{condition} becomes $\mathbf{d}_k\cdot\mathbf{d}'_k=0$, i.e.,
\begin{equation}\label{condition2}
  1+qq'-\eta\eta'+(q+q')\cos k=0.
\end{equation}
For $|1+qq'-\eta\eta'|<|q+q'|$, a critical interior mode $k_c=\arccos(-\frac{1+qq'-\eta\eta'}{q+q'})$ can be found. The interior mode is defined as $k\neq0$ and $k\neq\pi$ in Ref. \cite{R70}

$\mathcal{PT}$-symmetric phase consists two separate regions, ($\alpha$) and ($\beta$) (see Fig. \ref{figure1}). We first consider the quench protocol from ($\alpha$) to ($\beta$). Taking $\eta=\eta'=0.4$, $q=0.5$ and $q'=2$ as an example, we plot Fisher zeros and the rate function of the Loschmidt echo in Fig. \ref{figure2}. It can be seen that the line of Fisher zeros cuts the $\mathrm{Im}(z)$ axis [see Fig. \ref{figure2}(a1)], giving rise to nonanalytic behavior (cusp singularity) of $r(t)$ [see the red curve in Fig. \ref{figure2}(a2)], which implies that DQPT is occurring.
We also plot the dynamical topological order parameter--winding number
\begin{equation}\label{}
  \nu(t)=\frac{1}{2\pi}\int_0^\pi\frac{\partial\Phi_g(k,t)}{\partial k}dk,
\end{equation}
where $\Phi_g(k,t)$ is the Pancharatnam phase for mode $k$. The occurrence of DQPT is always accompanied by integer jumps of the winding number at the critical times [see the blue curve in Fig. \ref{figure2}(a2)]. Conversely, a DQPT also occurs when quenching from region ($\beta$) to region ($\alpha$), with the corresponding topological order parameter likewise exhibiting integer jumps at the critical times.  

In contrast, if the quench is performed with region ($\alpha$) or ($\beta$)--for instance, with $\eta=\eta'=0.4$, $q=1.5$ and $q'=2$ as an example--the line of Fisher zeros does not cut the $\mathrm{Im}(z)$ axis, and hence no DQPT can occur [see Fig. \ref{figure2}(b1) and (b2)]. To sum up, within the $\mathcal{PT}$-symmetric phase, only quenches between regions I and II can induce a DQPT, during which the topological order parameter undergoes integer jumps at the critical times.

It should be noted that the rate function can be negative, implying that during the dynamics of a non-Hermitian system, the absolute value of the return probability exceeds one, which is a distinctive feature of the biorthogonal theoretical framework.

\begin{figure}
\begin{center}
\includegraphics[width=8cm]{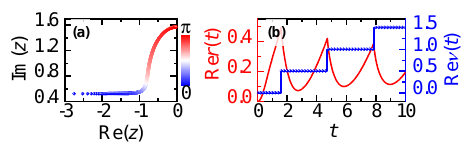}
\caption{(Color online) Lines of Fisher zeros (a), the time evolution of the real parts of rate function $\mathrm{Re}r(t)$ [red curves in (b)], the winding number $\mathrm{Re}\nu(t)$ [blue curves in (b)] for the quench of $q$ from $q=1$ to $q'=2$. The other parameters are $\eta=\eta'=0$ and $l=0$.}
\label{figure3}
\end{center}
\end{figure}

For $|1+qq'-\eta\eta'|=|q+q'|$, a critical boundary mode $k_c=0$ or $\pi$ can be found. The boundary mode is defiend as $k=0$ or $\pi$ in Ref. \cite{R70}. In this case, we find that a quench from the critical point between $\mathcal{PT}$-($\alpha$) and $\mathcal{PT}$-($\beta$), i.e. $(q=1,\eta=0)$, to either the $\mathcal{PT}$ phase ($\alpha$) or ($\beta$) gives rise to a new DQPT. Taking $\eta=\eta'=0$, $q=1$ and $q'=2$ as an example, we plot Fisher zeros, the rate function, and the winding number in Fig. \ref{figure3}. We observe that the Fisher zeros coalesce into a continuous curve that touches $\mathrm{Im}(z)$ axis at a critical boundary mode $k_c=\pi$ [see Fig. \ref{figure3}(a)], giving rise to nonanalytic behavior (cusp singularity) of the rate function of the Loschmidt echo which implies DQPT occurring [see the red curves in Fig. \ref{figure3}(b)]. It is worth noting that, unlike the previous DQPT with integer jumps in the winding number, the DQPT here is accompanied by winding numbers with half-integer jumps.

\subsection{DQPT for the quench from the $\mathcal{PT}$-symmetry-broken phase to the $\mathcal{PT}$-symmetric phase}
In this section, we examine the quench from the $\mathcal{PT}$-symmetry-broken phase, including I and II, to the $\mathcal{PT}$-symmetric phase. 

\begin{figure}
\begin{center}
\includegraphics[width=8cm]{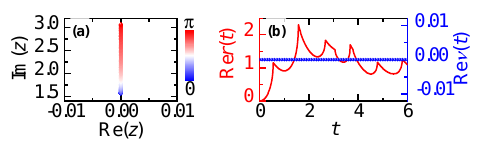}
\caption{(Color online) Lines of Fisher zeros (a), the time evolution of the real parts of rate function $\mathrm{Re}r(t)$ [red curves in (b)], the winding number $\mathrm{Re}\nu(t)$ [blue curves in (b)] for the quench of $\eta$ from $\eta=2$ to $\eta'=0.2$. The other parameters are $q=q'=0.5$ and $l=0$. }
\label{figure4}
\end{center}
\end{figure}

First, we consider the quench from $\mathcal{PT}$-symmetry-broken phase II to the $\mathcal{PT}$-symmetric phase. In this case, all modes from $k=0$ to $k=\pi$ satisfy the DQPT condition given in Eq. \eqref{condition}, indicating that all the Fisher zeros are located on $\mathrm{Im}(z)$ axis [see Fig. \ref{figure4}(a)], implying a very anomalous DQPT that the critical time is no longer periodic [see the red curves in Fig. \ref{figure4}(b)]. Since all modes are involved in DQPT, the winding number becomes ill-defined. Consequently, our calculations show that the winding number is invariably zero and thereby exhibits no correspondence to the DQPT behavior [see the blue curves in Fig. \ref{figure4}(b)].

\begin{figure}
\begin{center}
\includegraphics[width=8cm]{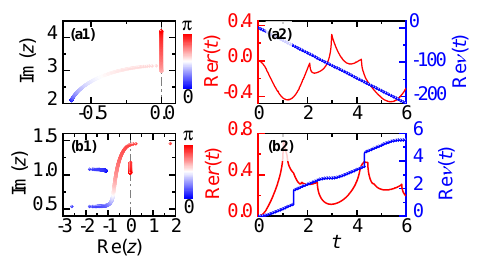}
\caption{(Color online) Lines of Fisher zeros (a1-b1), the time evolution of the real parts of the rate function $\mathrm{Re}r(t)$ [red curves in (a2) and (b2)] and the winding number $\mathrm{Re}\nu(t)$ [blue curves in (b1) and (b2)] for quenches $\eta=1\rightarrow \eta'=0$ with $q=q'=0.5$ (a1-a2) and $q=0.9\rightarrow q'=2$ with $\eta=\eta'=0.4$ (b1-b2). The Fisher zeros are calculated by considering $l=0$. }
\label{figure5}
\end{center}
\end{figure}

Then we investigate the quench from $\mathcal{PT}$-symmetry-broken phase I to $\mathcal{PT}$-symmetric phase. In the $\mathcal{PT}$-symmetry-broken phase I, the momentum-space modes can be classified into two types: The modes satisfy $|\eta|>|1+qe^{ik}|$, i.e., $\arccos\frac{\eta^2-1-q^2}{2q}<k\leq\pi$, which have purely imaginary energies; whereas those with $|\eta|<|1+qe^{ik}|$, i.e., $0\leq k<\arccos\frac{\eta^2-1-q^2}{2q}$, possess real energies. Accordingly, the Fisher zeros also separate into two sets: One associated with the purely imaginary energy modes, and the other with the real energy modes. As discussed previously for the $\mathcal{PT}$-antisymmetric phase, the Fisher zeros corresponding to purely imaginary energy modes ($ \arccos\frac{\eta^2-1-q^2}{2q} < k \leq \pi $) all lie on the $\mathrm{Im}(z)$ axis [see Fig. \ref{figure5}(a1) and (b1)]. These modes trigger a DQPT whose critical time is no longer periodic [see the red curves in Fig. \ref{figure5}(a2)]. Although the winding number evolves over time, its dynamic behavior does not reflect this DQPT [see the blue curves in Fig. \ref{figure5}(a2)]. For real-energy modes ($0 \leq k < \arccos\frac{\eta^2-1-q^2}{2q}$), if Eq. \eqref{condition2} is satisfied, the line of the corresponding Fisher zeros cuts the $\mathrm{Im}(z)$ axis [see Fig. \ref{figure5}(b1)]. In this case, an additional DQPT with periodic critical time emerges, characterized by the cusp singularity of the rate function at the periodic critical times. At these periodic critical times, the winding number exhibits integer jumps [blue curves in Fig. \ref{figure5}(b2)].

It is noteworthy that the dynamical phase transitions identified here---integer jumps driven by critical interior modes, half-integer jumps driven by critical boundary modes, and anomalous transitions with all modes on the imaginary axis beyond winding-number characterization---have recently been classified in the Hermitian one-dimensional XY model \cite{R70}. This suggests that the phenomena of integer, half-integer, and winding-number-undefined DQPTs are universal features of one-dimensional two-band systems, transcending the Hermitian versus non-Hermitian distinction.

\section{Conclusions}\label{Conclusion}

In this paper, we have refined the biorthogonal framework, leading to a gauge-invariant non-Hermitian quantum theory in which both left and right vectors satisfy the Schr\"{o}dinger equation. This gauge-invariant non-Hermitian quantum theory naturally reduces to standard quantum mechanics in the Hermitian limit and encompasses the open-system effective non-Hermitian evolution as a special case. As a concrete application, we analyzed the dynamical phase transition in a one-dimensional Su-Schrieffer-Heeger (SSH) model within this framework. In particular, using our formulation, the orthogonality condition for dynamical quantum phase transitions in Hermitian two-band systems, namely $\mathbf{d}_{k}^i\cdot\mathbf{d}_{k}^f=0$, is naturally generalized to the non-Hermitian case, taking the form $\mathrm{Re}\Bigl[\frac{\mathbf{d}_{k}^i}{d_{k}^i}\cdot\frac{\mathbf{d}_{k}^f}{d_{k}^f}\Bigr]=0$. In addition to dynamical quantum phase transitions that can be described by winding numbers, we have also discovered entirely new ones that cannot be characterized by winding numbers. We hope that this gauge-invariant non-Hermitian quantum theory will find broad applications in the study of non-Hermitian systems.

\section*{Acknowledgement}
This work was supported by the National Natural Science Foundation of China (Grant No. 11705099) and the Talent Introduction Project of Dezhou University of China (Grant Nos. 2020xjpy03 and 2019xgrc38).

\appendix
\section*{Derivation of dynamic phase transition conditions}
\label{APDIX:Derivation}

 For the initial ground state of non-Hermitian SSH model, the free energy density (\ref{free energy density}) can be calculated analytically, yielding
\begin{widetext}
\begin{equation}\label{}
  f(z)=-\frac{1}{2\pi}\int_{k\in BZ}dk
  \ln\Biggl[\frac{e^{-zd_k^f}}{2}\biggl(1-\frac{\mathbf{d}_k^i}{d_k^i}\cdot\frac{\mathbf{d}_k^f}{d_k^f}\biggr)
  +\frac{e^{zd_k^f}}{2}\biggl(1+\frac{\mathbf{d}_k^i}{d_k^i}\cdot\frac{\mathbf{d}_k^f}{d_k^f}\biggr)\Biggr].
\end{equation}
\end{widetext}
In the thermodynamic limit the zeros of the partition function in the complex plane coalesce to a family of lines labeled by a number $l\in \mathbb{Z}$
\begin{equation}\label{Zeros}
z_n(k)=\frac{1}{2d_{k}^f}\Biggl[\ln\Bigg|\frac{1-\frac{\mathbf{d}_k^i}{d_k^i}\cdot\frac{\mathbf{d}_k^f}{d_k^f}}
{1+\frac{\mathbf{d}_k^i}{d_k^i}\cdot\frac{\mathbf{d}_k^f}{d_k^f}}\Bigg|+i\Theta+i(2l+1)\pi\Biggr],
\end{equation}
where 
\begin{equation}
 \Theta=\arg\Biggl[\frac{1-\frac{\mathbf{d}_k^i}{d_k^i}\cdot\frac{\mathbf{d}_k^f}{d_k^f}}{1+\frac{\mathbf{d}_k^i}{d_k^i}\cdot\frac{\mathbf{d}_k^f}{d_k^f}}\Biggr]    
\end{equation}
is the argument principal value. From Eq. \eqref{Zeros}, we can see that for $d_{k}^f$ is real, if 
\begin{equation}
    \Bigg|\frac{1-\frac{\mathbf{d}_k^i}{d_k^i}\cdot\frac{\mathbf{d}_k^f}{d_k^f}}
{1+\frac{\mathbf{d}_k^i}{d_k^i}\cdot\frac{\mathbf{d}_k^f}{d_k^f}}\Bigg|=1,
\end{equation}
a DQPT occurs at the critical times
\begin{equation}\label{}
  t_{c,l}=t_\theta+t_c\Bigl(l+\frac{1}{2}\Bigr),~~l=0,1,2\cdots
\end{equation}
with $t_\theta=\Theta/(2d_k^f)$, $t_c=\pi/d_{k}^f$. The mode satisfies these conditions is called the critical mode $k_c$, which is determined by 
\begin{equation}\label{}
  \frac{1-\frac{\mathbf{d}_k^i}{d_k^i}\cdot\frac{\mathbf{d}_k^f}{d_k^f}}{1+\frac{\mathbf{d}_k^i}{d_k^i}\cdot\frac{\mathbf{d}_k^f}{d_k^f}}=e^{i\Theta}.
\end{equation}
This implies
\begin{equation}\label{}
  \frac{1-\Bigl(\frac{\mathbf{d}_k^i}{d_k^i}\cdot\frac{\mathbf{d}_k^f}{d_k^f}\Bigr)^\ast}{1+\Bigl(\frac{\mathbf{d}_k^i}{d_k^i}\cdot\frac{\mathbf{d}_k^f}{d_k^f}\Bigr)^\ast}
=\frac{1+\frac{\mathbf{d}_k^i}{d_k^i}\cdot\frac{\mathbf{d}_k^f}{d_k^f}}{1-\frac{\mathbf{d}_k^i}{d_k^i}\cdot\frac{\mathbf{d}_k^f}{d_k^f}},
\end{equation}
which leads to
\begin{equation}\label{}
  \Biggl(\frac{\mathbf{d}_k^i}{d_k^i}\cdot\frac{\mathbf{d}_k^f}{d_k^f}\Biggr)^\ast=-\frac{\mathbf{d}_k^i}{d_k^i}\cdot\frac{\mathbf{d}_k^f}{d_k^f}.
\end{equation}
As a result, the condition of dynamical phase transition becomes
\begin{equation}\label{conditiona}
  \mathrm{Re}\Biggl[\frac{\mathbf{d}_k^i}{d_k^i}\cdot\frac{\mathbf{d}_k^f}{d_k^f}\Biggr]=0.
\end{equation}
The condition of the dynamical phase transition in the Hermitian system, namely $\mathbf{d}_k^i\cdot\mathbf{d}_k^f=0$, is the special case of Eq. (\ref{conditiona}).

\end{document}